\documentclass[12pt,onecolumn]{IEEEtran}
\ifCLASSINFOpdf
\else
\fi

\newtheorem{theorem}{Theorem}
\newtheorem{corollary}{Corollary}
\newtheorem{lemma}{Lemma}
\newtheorem{remark}{Remark}

\newtheorem{definition}{Definition}
\newtheorem{example}{Example}

\newtheorem{proposition}{Proposition}

\usepackage[top=0.8in, bottom=0.8in, left=0.8in, right=0.8in]{geometry}
\usepackage{epstopdf, multicol}
\usepackage{amsmath,amsfonts, amssymb,color, mathrsfs}
\usepackage{tikz, subcaption}
\usetikzlibrary{shapes.geometric}
\usepackage[font={small,it}]{caption}
\usepackage{graphicx, enumitem}
\usepackage{amsmath, cite}
\usepackage{amsfonts}
\usepackage{graphicx}
 \usepackage{algorithm} 
 \usepackage{algorithmic} 

\begin{document}
%
\title{The Game-Theoretic Formation of Interconnections Between  Networks}
%
%
%
\date{}
\author{Ebrahim Moradi Shahrivar~and~Shreyas Sundaram
\IEEEcompsocitemizethanks{\IEEEcompsocthanksitem Ebrahim Moradi Shahrivar is with the Department of Mechanical and Mechatronics Engineering at the University of Waterloo. E-mail: {\tt emoradis@uwaterloo.ca}.
\IEEEcompsocthanksitem Shreyas Sundaram is with the School of Electrical and Computer Engineering at Purdue University.  E-mail: {\tt sundara2@purdue.edu}.}

\thanks{This material is based upon work supported by the Natural Sciences and Engineering Research Council of Canada.}
}

%
%

\markboth{}%
{Shell \MakeLowercase{\textit{et al.}}: Bare Demo of IEEEtran.cls for IEEE Journals}
%



\maketitle


\begin{abstract}
We introduce a network design game where the objective of the players is to design the interconnections between the nodes of two different networks $G_1$ and $G_2$ in order to maximize certain local utility functions. In this setting, each player is associated with a node in $G_1$ and has functional dependencies on certain nodes in $G_2$.  We use a distance-based utility for the players in which the goal of each player is to purchase a set of edges (incident to its associated node) such that the sum of the distances between its associated node and the nodes it depends on in $G_2$ is minimized. We consider a heterogeneous set of players (i.e., players have their own costs and benefits for constructing edges). We show that finding a best response of a player in this game is NP-hard. Despite this, we characterize some  properties of the best response actions which are helpful in determining a Nash equilibrium for certain instances of this game. In particular, we prove existence of pure Nash equilibria in this game when $G_2$ contains a star subgraph, and provide an algorithm that outputs such an equilibrium for any set of players. Finally, we show that the price of anarchy in this game can be arbitrarily large.
\end{abstract}


%
\IEEEpeerreviewmaketitle

\section{Introduction}
\label{sec:intro}
There is a growing realization that many large scale networks consist of interconnected subnetworks \cite{P17,  Radicchi, porter}. Examples include coupled energy infrastructure and communication networks \cite{yazdani},    cyber-physical systems \cite{P15}, and transportation networks (such as the flight networks of different airlines) \cite{ATN}. 
Understanding the structure of large-scale networks and the implications of this structure for the effective functioning of the network has been the subject of many studies throughout the past decade \cite{P3, goyal}. One approach to investigate this problem is through the framework of {\it random graphs} where each subnetwork is drawn from a certain probability distribution \cite{P1, P2}. For (single) random networks, properties such as connectivity, robustness against structural and dynamical failures, and edge expansion have been widely investigated in the literature \cite{hogan2,coherence,Ghaderi13,P7}, with recent extensions to interconnected networks  \cite{yagan, yazdani, P12, P16, Shahrivar15}. 

An alternative perspective on understanding the structure of networks is to view the edges as being optimally placed (either by a central designer, or by different decision makers) in order to maximize some given utility function(s) \cite{P8, P9, P10, basar}.  The classical literature on optimal network design has predominantly focused on the construction of a single network \cite{P3, P4}. In \cite{MoradiShahrivar13}, we proposed a multi-layer network formation setting in which, given a network $G=(V, E)$, the network designer aims to find a network $G_1=(V, E_1)$ such that distances between nodes that are neighbors in $G$ are minimized in $G_1$; in this context, networks $G$ and $G_1$ represent different types of relationships between the set of nodes $V$. We then exploited this setting to formulate a multi-layer network formation game where each layer corresponds to a player that is optimally choosing its edge set in response to the edge sets of the other players.

In this paper, we consider the game-theoretic formation of edges {\it between} two given networks $G_1=(V_1, E_1)$ and $G_2=(V_2, E_2)$ on two different sets of nodes $V_1$ and $V_2$.  We assume that there are dependencies between nodes in $V_1$ and $V_2$, i.e., some of the nodes in $V_1$ require connections to (or information from) some of the nodes in $V_2$ in order to function.  These dependencies are captured by a bipartite network $G_I=(V_1 \cup V_2, E_I)$ where $E_I \subseteq V_1 \times V_2$, and an edge $(v_i,v_j) \in E_I$ indicates that $v_i\in V_1$ is dependent on $v_j\in V_2$.  We assume that each node in  $V_1$ is a player and builds a set of edges between itself and nodes in $V_2$ in order to maximize a distance-based utility function. As a motivating abstraction for this problem, consider a cyber-physical system where $G_1$ is a power network (with the nodes representing substations) and $G_2$ is a sensor network. Suppose that the sensor nodes are responsible for gathering critical information (e.g., power usage, line failures, etc.) from different geographical regions and this information is required by the power stations that supply electricity to those regions. This setting has been investigated from different perspectives over the past few years \cite{marzieh,cyberpower, yagan, P15, P17, yazdani,buldyrev2010} and our results add to this literature by studying the game-theoretic allocation of interconnecting edges between the power and sensor networks. We model dependencies between the substations and sensors (which correspond to nodes in $G_1$ and $G_2$, respectively) by the network $G_I$. Suppose that neighboring nodes in each network are capable of exchanging information with each other. The substation operators wish to construct connections to the sensor network in such a way that they minimize the number of hops required to gather data from their interdependent nodes (where the number of hops is measured with respect to the connections within $G_1$ and $G_2$ and the edges constructed between the networks). This leads to an {\it interconnection network design game} (INDG) with distance utilities where the utility of each player (operator) depends on its own set of edges as well as the set of edges constructed by other players. 

The INDG setting also matches the framework studied in \cite{Moskvina2016} for merging two social networks where the goal is to construct a set of edges between the networks such that the integrated network has diameter no more than a fixed value. Besides considering a cost for constructing edges and having a different utility function, the {\it nodes} are the decision makers in our setting, whereas \cite{Moskvina2016} assumes a central network designer.
 Distance-based utilities have also been used to study computer networks (where nodes represent the computers and edges are the communication links) \cite{P6, TNSE}. In this case, network $G_I$ models the virtual dependencies among the computers in cluster $G_1$ and cluster $G_2$, indicating the set of pairs of nodes that wish to exchange information. The designed interconnection network represents the physical communication network between the two clusters. Yet another application of the INDG with distance utilities arises in studying interconnections between the transportation networks of two countries. We will elaborate on this example in Section \ref{dist_game}.

We start our investigation of the INDG by showing that it is NP-hard to find a best response for each player. Despite the NP-hardness of the problem, we characterize some useful properties of the best response  which consequently enable us to determine a Nash equilibrium instance for certain cases of the game.  
Specifically, we study the existence of Nash equilibria in an INDG with distance utilities when network $G_2$ has a star subgraph (similar to the ``hub-and-spoke'' structure seen in various transportation networks \cite{jeanpaul, USPS} or in sensor networks with fusion centers \cite{sensorwireless,centralSN}) and there is full interdependency between nodes in $G_1$ and $G_2$. We show that this setting possesses a Nash equilibrium for any set of players with arbitrary benefit functions and edge costs.  We partition the set of players into two sets consisting of high and low edge cost players and show that in any Nash equilibrium, all of the high-cost players that have a low-cost player in their vicinity ``free ride'' and choose not to construct any interconnections to $G_2$.  At the end, we provide some insights about our results via a simulation involving random network models that have been previously used to capture interdependencies between power and sensor/communication networks \cite{cyberpower,buldyrev2010,yazdani,P16}. Our simulations suggest that the social welfare of the constructed networks is higher when all of the players have equal cost of constructing edges, compared to the case where they have heterogeneous edge costs.
 
\section{Definitions}
\label{sec:defs}
An undirected network (or graph) is denoted by $G=(V,E)$ where $V=\{ v_1, v_2, \dots , v_n\}$ is the set of nodes (or vertices) and $E \subseteq \{(v_i,v_j) \lvert v_i,v_j \in V, v_i \neq v_j \}$ denotes the set of edges. If there is an edge between two nodes, they are said to be neighbors. The number of neighbors of a node $v_i \in V$ in graph $G$ is called its degree and is denoted by $\deg_i(G)$. 
A path from node $v_1$ to $v_k$ in graph $G$ is a sequence of distinct nodes $v_1 v_2 \cdots v_k$ where there is an edge between each pair of consecutive nodes of the sequence. The length of a path is the number of edges in the sequence. We denote the shortest distance between nodes $v_i$ and $v_j$ in graph $G$ by $d_G (v_i,v_j)$. If there is no path from $v_i$ to $v_j$, we take $d_G(v_i,v_j)=\infty$. The diameter of the graph $G$ is $\max_{v_i,v_j \in V, v_i \ne v_j}d_G(v_i,v_j)$.
A cycle is a path of length two or more from a node to itself. A graph $G'= (V', E')$ is called a subgraph of $G=(V,E)$, denoted as $G' \subseteq G$, if $V' \subseteq V$ and $E' \subseteq E \cap \{ V' \times V' \}$. A graph is connected if there is a path from every node to every other node.   A subgraph $G' = (V', E')$ of $G$ is a component if $G'$ is connected and there are no edges in $G$ between nodes in $V'$ and nodes in $V \setminus V'$.  
A graph $G=(V, E)$ is called bipartite if there exist two {\it disjoint} subsets $V_1, V_2 \subseteq V$ such that $V_1 \cup V_2=V$ and $E \subseteq V_1 \times V_2$, i.e., $G$ does not have any edge with both endpoints in $V_1$ or $V_2$. The set of all possible bipartite graphs with two partitions $V_1$ and $V_2$ is denoted by $G^{V_1 \times V_2}$.
\section{Distance-Based Utility}
Jackson and Wolinsky introduced a canonical problem in network formation which involves distance-based utilities \cite{P3}. In their formulation, each node is a decision maker, and chooses its connections to other nodes in the network. In any formed network, each node receives a benefit of  $b(k)$ from nodes that are $k$ hops away, where $b:\{1,2,\cdots,n-1,\infty\} \to \mathbb{R}_{\geq 0}$ is a real-valued, nonincreasing, nonnegative function (i.e., nodes that are further away provide smaller benefits) and $b(\infty)=0$. Furthermore, constructing the edge $(v_i, v_j)$ incurs a cost of $c$ to both endpoints $v_i$ and $v_j$.  
The total utility that node $v_i$ receives from the constructed network $G=(V, E)$ is
\begin{equation}\label{dist_distance_util}
u_i(G)=\left( \sum_{ v_j \in V: v_i \ne v_j} b(d_G(v_i,v_j))\right) - c~\text{deg}_i (G).
\end{equation}

Thus, the nodes have to compromise between adding  more links (which provides a larger benefit by reducing the distances between nodes) and decreasing the cost by using fewer edges.  
When $b(\cdot)$ is a strictly decreasing function, there are only a few different kinds of socially optimal (or efficient) networks, depending on the relative values of the link costs and  connection benefits: the empty network (for high edge costs), star (for medium edge costs) and the complete network (for low edge costs) \cite{P4}.

\section{Interconnection Network Design Game}\label{dist_game}
Assume that we are given two arbitrary networks $G_1=(V_1, E_1)$ and $G_2=(V_2, E_2)$.  
In this paper, we consider a setting in which each node in $V_1$ constructs a set of edges to nodes in $V_2$ such that some utility function is maximized. This leads to a game with the nodes of $G_1$ as the players. 

\begin{definition} \label{distgame}
Consider two arbitrary networks $G_1=(V_1, E_1)$ and $G_2=(V_2, E_2)$ with $V_1=\{x_{1}, \cdots, x_{n}\}$ and $V_{2}=\{y_{1}, \cdots, y_{m}\}$.  
An instance of the {\it interconnection network design game} (INDG) $\mathcal{G}=(P, (S_i)_{P_i \in P}, (\Psi_i)_{P_i \in P}, G_1, G_2)$ has a set of $n$ players $P=\{P_1, P_2, \cdots, P_{n}\}$ where player $P_i$ is  associated with node $x_i \in V_1$ for $1 \le i \le n$. The strategy space of player $P_i$ is $S_i = 2^{\{x_i\} \times V_2}$, i.e., all possible subsets of edges from $x_i$ to nodes in $V_2$. The action of player $P_i$ is an element of $S_i$ and is denoted by $W_i$, i.e., $W_i$ is a set of edges from $x_i$ to a certain subset of $V_2$.  By an abuse of notation, we take $B=\cup_{j =1}^{n}W_j$ to indicate the bipartite graph $B = (V_1 \cup V_2, \cup_{j=1}^{n}W_j)$. 
The utility of player $P_i$ is given by a function $\Psi_i: S_1 \times S_{2} \times \dots \times S_{n} \to \mathbb{R}$, where the $j^{th}$ argument\footnote{The utility function $\Psi_i$ is also a function of $G_1$ and $G_2$ which will be omitted from the argument list as long as it is clear from the context.} is the action of the $j^{th}$ player for $1 \leq j \leq n$. 
\end{definition}

The characteristics of the game and the optimal strategies for each player will depend on the form of the utility functions $\Psi_i$. In this paper, we consider a modified version of the distance utility function in \eqref{dist_distance_util} as the payoff to the players. Specifically, we assume that there are dependencies between nodes in the graphs $G_1$ and $G_2$  which is represented by a bipartite network $G_I=(V_1 \cup V_2, E_I)$ with two partitions $V_1$ and $V_2$  and $E_I \subseteq V_1 \times V_2$. Let $I_i \subseteq V_2,~1 \le i \le n$, denote the set of neighbors of $x_i \in V_1$ in the network $G_I$. Then the objective of player $P_i$ is  to find the optimal set of edges to construct to $V_2$ such that distance between its associated node $x_i$ and the set of nodes in $I_i$ is minimized.   
In addition to the technological applications that we mentioned in Section~\ref{sec:intro}, the INDG can be utilized to model problems in transportation. For instance consider a modified version of the problem studied in \cite{P12} where we are given the traffic flow between cities of two different countries $C_1$ and $C_2$. Each of these countries has a domestic transportation service which connects its cities and is modeled by networks $G_1$ and $G_2$. A city in $C_1$ and a city in $C_2$ are said to be interdependent if the traffic flow between them is higher than some threshold, and this interdependency is represented by an edge between them in the network $G_I$. The players of the game correspond to transportation service planners at each node in $C_1$, who are faced with the problem of finding the optimal set of routes to establish from their associated city to cities of the country $C_2$ such that distance between the interdependent cities is minimized.     
It is clear that the structure of the interconnection between cities inside the countries $C_1$ and $C_2$ (modeled as $G_1$ and $G_2$) affects the optimal decisions made by the players.

\begin{definition}
An instance 
$$
\mathcal{G}=(P, (S_i)_{P_i \in P}, (u_i)_{P_i \in P}, G_1, G_2, G_I)
$$
of the game in Definition \ref{distgame} is said to be an {\it interconnection network design game with distance utilities} if the utility function of player $P_i$, $1 \le i \le n$, with action $W_i \in S_i$ has the form
\begin{align} \label{distgame_util}
\Psi_i(W_1, \cdots, W_{n})&= u_i(\cup_{j =1}^{n}W_j|G_1, G_2, G_I)\\ \nonumber
&= \left( \sum_{y \in I_i} b_i(d_{G}(x_i,y))\right) -c_i|W_i|,
\end{align}
where $G=(V_1 \cup V_2, E_1 \cup E_2 \cup (\cup_{j=1}^{n}W_j))$.
\label{def:INDG_DU} 
\end{definition}

As we can see in the utility function $u_i(\cdot)$, only the distances between node $x_i$ and the set of nodes $I_i$ matter. Furthermore, each player has to pay only for his/her constructed edges. The benefit functions $b_i(\cdot)$ are nonnegative, nonincreasing and satisfy $b_i(\infty) = 0$, and all costs $c_i$ are positive, and can be different across players. 

We will use $W_{-i}$ to denote the vector of actions of all players except player $P_i$, and use $\Psi_i(W_i,W_{-i})$ to denote the utility of player $P_i$ with respect to the given vector $(W_1, W_2, \ldots, W_{n})$.  Based on the definition of the game, we say that a vector of actions $(W_1, W_2, \ldots, W_{n})$ is a {\it Nash equilibrium} if and only if $W_i\in\operatorname*{arg\,max}_{W \in S_i} \Psi_i(W, W_{-i})$ for all $i \in \{1, 2, \ldots, n\}$.  In this case, $W_i$ is said to be a {\it best response} action to $W_{-i}$ with respect to the utility function $\Psi_i$. 
For the rest of this paper, whenever we say INDG, by default we mean an interconnection network design game with distance utilities.

\begin{remark}
The benefit function $b_i(\cdot)$ can be chosen to capture how quickly (in terms of number of hops) node $v_i \in V_1$ needs to communicate with its interdependent nodes in $V_2$.  For example, consider again the cyber-physical system abstraction described in Section~\ref{sec:intro}, where the nodes in $V_1$ are power substations and nodes in $V_2$ are sensors that  measure certain quantities of interest.  If substation $v_i \in V$ is able to tolerate a routing delay of up to $k$ hops from each of the sensors it depends on, but higher routing delays are useless, then the associated benefit function can be chosen as $b_i(1) = \cdots = b_i(k) > 0$, and $b_i(d) = 0$ for $d > k$.  Alternatively, if node $v_i$ is able to tolerate any routing delay, but would prefer shorter delays, $b_i(\cdot)$ can be chosen to be an appropriate strictly decreasing function.  Our formulation allows different nodes to have different benefit functions and edge costs, encoding heterogeneity in the players of the game. 
\end{remark}

\section{Characteristics of the Best Responses}
\label{sec:full_interdependent}
In this section, we characterize some important properties of the best response actions for the players. We start by determining the complexity of finding a best response action for the players in the INDG.

\subsection{Complexity}
\label{sec:complexity}
In order to characterize the complexity of finding the best response actions for the papers, we first formulate the {\it decision problem} corresponding to optimizing the utility \eqref{distgame_util}, as follows.\footnote{Decision problems are those with ``yes'' or ``no'' answers, and form the basis of the complexity classes P and NP.  Since optimization problems can be solved by repeatedly solving a corresponding decision problem (e.g., by determining whether there is a solution that provides a utility larger than a certain threshold), showing that the decision problem is NP-hard is sufficient to show NP-hardness of the optimization problem.  We refer to standard textbooks such as \cite{CLRS} for more details and background on complexity theory.}

\begin{definition}
	{\bf Best Response Interconnection (BRI).}\\
	\textbf{INSTANCE}: A given instance 
	$$
	\mathcal{G}=(P, (S_i)_{P_i \in P}, (u_i)_{P_i \in P}, G_1, G_2, G_I),
	$$
	of INDG, a player $P_j \in P$, a joint strategy by all other players $W_{-j}=\cup_{i\neq j} W_i$ and a threshold $r \in \mathbb{R}_{> 0}$.
	
	\textbf{QUESTION}: Does there exist an action $W_j \in S_j$ for the player $P_j$ such that 
	\begin{align*}
	u_j(W_j \cup W_{-j}&|G_1, G_2, G_I)= \left( \sum_{y \in I_j} b_j(d_{G}(x_j,y))\right) -c_j|W_j| \ge r,
	\end{align*}
	where $G=(V_1 \cup V_2, E_1 \cup E_2 \cup W_j \cup W_{-j})$?
\end{definition}

We now provide the following theorem showing that finding a best response for the players, given arbitrary networks $G_1, G_2, G_I$, and arbitrary non-increasing benefit functions $b_i(\cdot)$ and edge costs $c_i>0$ for the players, is impossible in polynomial-time (unless P = NP).

\begin{theorem}\label{thm:dis_nphard}
	The Best Response Interconnection problem is NP-hard.
\end{theorem}

To prove this theorem, we provide a reduction from the NP-complete Dominating Set Problem \cite{CLRS}. A dominating set of the network $G_d=(V_d, E_d)$ is a subset $S \subseteq V_d$ such that for all $u \in V_d \setminus S$, $u$ has a neighbor in the set $S$.

\begin{definition}
{\bf Dominating Set Problem.}\\
\textbf{INSTANCE}: Network $G_d=(V_d, E_d)$ and positive integer $k \le |V_d|$.\\
\textbf{QUESTION}: Does the network $G_d$ have a dominating set $S$ with $|S| \le k$?
\end{definition}

We are now in place to prove Theorem \ref{thm:dis_nphard}.

\begin{IEEEproof}[Proof of Theorem \ref{thm:dis_nphard}]	
Given an instance of the dominating set problem with $G_d=(V_d, E_d)$ and $k$, define an instance of the  BRI problem with $G_1=(V_1, E_1), G_2=(V_2, E_2)$ and $G_I=(V_1 \cup V_2, E_I)$ as follows
\begin{align} \nonumber
V_1=\{x_1\}, ~V_2&=V_d,~E_1=\phi, ~E_2=E_d,~E_I=V_1 \times V_2\\ \nonumber
b_1(3)&<b_1(1)-c_1<b_1(2)\\ \label{eqn:constructed_BRI}
r=k&(b_1(1)-c_1)+(|V_2|-k)b_1(2). 
\end{align}
For example $c_1=2, b_1(1)=4, b_1(2)=3, b_1(3)=1$ and $b_1(k)=0$ for all $k \ge 4$ satisfies the above conditions.
In the above instance of the BRI, there is only one node in $V_1$ (with associated player $P_1$), and this player is fully dependent on all nodes in $V_2$ (i.e., we have $I_1 = V_2$ in \eqref{distgame_util}).  Hence, the BRI problem is to determine whether $P_1$ has an action $W_1$ such that $u_1(W_1|G_1, G_2, G_I) \ge r$.

The above instance of the BRI problem can be constructed in polynomial time. In the rest of the proof, we show that the answer to the above instance of the BRI problem is ``yes'' if and only if the answer to the given instance of the Dominating Set Problem is ``yes".

Suppose that the graph $G_2=G_d$ has a dominating set $S \subset V_2$ with $|S|\le k$ and thus the answer to the given instance of the Dominating set problem is ``yes". Then by defining $W_1=\{(x_1,v)|v \in S\}$, the distance between node $x_1$ and any node in $V_2$ is at most 2. Since $|W_1| \le k$, 
\begin{align*}
u_1(W_1|G_1, G_2, G_I)&=|W_1|(b_1(1)-c_1)+(|V_2|-|W_1|)b_1(2) \\
&=|W_1|(b_1(1)-c_1)+(|V_2|-k)b_1(2)+(k-|W_1|)b_1(2)\\
&\ge |W_1|(b_1(1)-c_1)+(|V_2|-k)b_1(2)+(k-|W_1|)(b_1(1)-c_1)\\
&= r.
\end{align*}
Therefore, the answer to the constructed instance of the BRI problem in \eqref{eqn:constructed_BRI} is ``yes" as well.

Next suppose that the answer to the defined instance of BRI in \eqref{eqn:constructed_BRI} is ``yes", i.e., there exists a $W_1 \in S_1$ such that $u_1(W_1|G_1, G_2, G_I) \ge r$. If there is a node $v \in V_2$ such that $d_G(x_1,v) \ge 3$, we can add the edge $(x_1, v)$ to $W_1$; this would increase the benefit of the network by at least $b_1(1)-b_1(3)$ and incur a cost of $c_1$. Since $b_1(1)-b_1(3)>c_1$, this would increase the utility of $P_1$. Thus without loss of generality we can take the distance between node $x_1$ and any node in $V_2$  to be at most 2 under the constructed edge set $W_1$.

Consider the set of nodes $S \subseteq V_2$ that are incident to at least one edge in $W_1$, i.e., $S=\{v \in V_2 |  (x_1, v) \in W_1\}$. All of the nodes in $V_2 \setminus S$ are connected to at least one of the nodes in $S$ due to the assumption that the distance between any node in $V_2$ and node $x_1$ is at most 2. Thus $S$ is a dominating set of the network $G_2$. On the other hand, the assumption that $u_1(W_1|G_1, G_2, G_I) \ge r$ yields
\begin{align*}
0 &\le u_1(W_1|G_1, G_2, G_I)-r\\
&= |W_1|(b_1(1)-c_1)+(|V_2|-|W_1|)b_1(2) -r\\
&= (|W_1|-k)(b_1(1)-c_1)+(k-|W_1|)b_1(2)\\
&=(|W_1|-k)(b_1(1)-c_1-b_1(2)).
\end{align*}
Since $b_1(1)-c_1<b_1(2)$, we must have that $|W_1|\le k$. Hence, $|S|=|W_1| \le k$. This means that network $G_2$ has a dominating set of size less than $k$. Thus the answer to the given instance of the Dominating Set Problem is ``yes".
\end{IEEEproof}

Given that BRI is a NP-hard problem, finding best response actions in the INDG with distance utilities is nontrivial in general. 
In the next section, we provide some properties of the best response actions that will be helpful in characterizing the best responses of the players in certain cases.

\subsection{Properties of the Best Response}

\begin{lemma}\label{lem:number_of_edge}
Let $W_j$ be a best response to $W_{-j}$ in the INDG 
$$
\mathcal{G}=(P, (S_i)_{P_i \in P}, (u_i)_{P_i \in P}, G_1, G_2, G_I).
$$
Then we have that 
\begin{enumerate}
\item $|W_j| \le |I_j|$.
\item If $b_j(1)>b_j(2)$, then $|W_j| = |I_j|$ if and only if $W_j=\{(x_j, y)|y \in I_j\}$.
\end{enumerate}
\end{lemma}
\begin{IEEEproof}
Let $G=(V_1 \cup V_2, E_1 \cup E_2 \cup W_j \cup W_{-j})$. We use contradiction to prove the first statement. Assume that $|W_j| > |I_j|$, then
\begin{align*}
u_j(W_j \cup W_{-j}|G_1, G_2, G_I)&= \left( \sum_{y \in I_j} b_j(d_{G}(x_j,y))\right) -c_j|W_j| \\
	&\le |I_j|b_j(1)-c_j|W_j| \\
	&< |I_j|(b_j(1)-c_j)\\
	&=u_j(W'_j \cup W_{-j}|G_1, G_2, G_I),
	\end{align*}
where $W'_j=\{(x_j, y)|y \in I_j\}$. Thus $W_j$ is not a best response to $W_{-j}$ which is a contradiction to the assumption of the lemma. 

To prove the second statement, note that if $W_j=\{(x_j, y)|y \in I_j\}$, then $|I_j|=|W_j|$. Thus we only have to show that when $b_j(1)>b_j(2)$, if  $|I_j|=|W_j|$, then $W_j=\{(x_j, y)|y \in I_j\}$. Assume by way of contradiction that there exists $y^* \in I_j$ such that $(x_j, y^*) \notin W_j$. This means that $d_G(x_j,y^*) \ge 2$ and thus
\begin{align*}
u_j(W_j \cup W_{-j}|G_1, G_2, G_I)&< |I_j|b_j(1)-c_j|I_j|\\
&= u_j(W'_j \cup W_{-j}|G_1, G_2, G_I),
\end{align*}
where, again, $W'_j=\{(x_j, y)|y \in I_j\}$.  This is a contradiction and thus we must have $\{(x_j, y)|y \in I_j\} \subseteq W_j$. We also know that $|W_j| \le |I_j|$ and therefore, have the required result.
\end{IEEEproof}

The next lemma characterizes a best response action of the players when the cost of constructing edges is less than a certain threshold. The proof follows the same reasoning as the proof in \cite{P4} for the formation of (single) networks under low edge costs.

\begin{lemma}\label{lem:lowlowcost}
Let $W_j$ be a best response to $W_{-j}$ in the INDG 
$$
\mathcal{G}=(P, (S_i)_{P_i \in P}, (u_i)_{P_i \in P}, G_1, G_2, G_I).
$$
If $c_j<b_j(1)-b_j(2)$, then $W_j=\{(x_j, y)|y \in I_j\}$. Furthermore, if $c_j=b_j(1)-b_j(2)$, then $W_j=\{(x_j, y)|y \in I_j\}$ is a best response action for player $P_j$.
\end{lemma}

\begin{IEEEproof}
Suppose that $y^* \in I_j$ and $(x_j,y^*) \notin W_j$. Then $b_j(d_G(x_j,y^*)) \le b_j(2)$ where $G=(V_1 \cup V_2, E_1 \cup E_2 \cup W_j \cup W_{-j})$. Adding the edge $(x_j,y^*)$ to $W_j$ increases the utility of $W_j$ by at least $b_j(1)-c_j-b_j(2)>0$ which contradicts the assumption that $W_j$ is a best response and thus $(x_j,y^*) \in W_j$. Hence $\{(x_j, y)|y \in I_j\} \subseteq W_j$. By Lemma \ref{lem:number_of_edge}, we know that $|W_j| \le |I_j|$ and therefore, $W_j=\{(x_j, y)|y \in I_j\}$. 

For the case that $c_j=b_j(1)-b_j(2)$, note that adding the edge $(x_j,y^*)$ to $W_j$ does not decrease the utility of $W_j$ and thus as in the above argument, $W_j=\{(x_j, y)|y \in I_j\}$ is a best response action for $P_j$.
\end{IEEEproof}

The next result gives an upper-bound on the maximum number of edges that a player $P_j$ with $ c_j > b_j(1)-b_j(2)$ will form in a Nash equilibrium.

\begin{lemma}\label{lem:highcost_action_changed}
Let $W_j$ be a best response to $W_{-j}$ in the INDG 
$$
\mathcal{G}=(P, (S_i)_{P_i \in P}, (u_i)_{P_i \in P}, G_1, G_2, G_I).
$$
If $b_j(1)-b_j(2) < c_j$, then $|W_j| \le |D|$, where $D$ denotes the  smallest dominating set of the network $G_2$. 
\end{lemma}
\begin{IEEEproof}
If $|I_j| \le |D|$, we have the result by the first statement of Lemma \ref{lem:number_of_edge}. Thus consider the case that $|I_j|>|D|$.
	Assume by way of contradiction that $|W_j| > |D|$. Let $G=(V_1 \cup V_2, E_1 \cup E_2 \cup W_j \cup W_{-j})$. Then 
	\begin{align*} 
    u_j( W_{j} \cup W_{-j}|G_1, G_2, G_I)&\le |W_j|(b_j(1)-c_j)+(|I_j|-|W_j|)b_j(2)\\
	&=|D|(b_j(1)-c_j)+(|W_j|-|D|)(b_j(1)-c_j)+(|I_j|-|W_j|)b_j(2)\\
	&< |D|(b_j(1)-c_j)+(|I_j|-|D|)b_j(2)\\
	&=u_j( W'_{j} \cup W_{-j}|G_1, G_2, G_I),\nonumber
	\end{align*}
	where $W'_j=\{(x_j, y)|y \in D\}$. Thus  $W'_j$ produces more utility than $W_j$ for player $P_j$ which is a contradiction to the assumption that $W_j$ is a best response to $W_{-j}$. 	
\end{IEEEproof}

We will apply Lemma \ref{lem:highcost_action_changed} later in Section \ref{sec:dist_nash} to determine a Nash equilibrium instance of the INDG when $G_2$ has a star subgraph. 
The next lemma provides a threshold on the edge costs of the players in order for them to have nonempty actions.

\begin{lemma}\label{lem:threshold}
Let $W_j$ be a best response to $W_{-j}$ in the INDG  
$$
\mathcal{G}=(P, (S_i)_{P_i \in P}, (u_i)_{P_i \in P}, G_1, G_2, G_I).
$$
If $c_j>b_j(1)+(|I_j|-1)b_j(2)$, then $W_j=\phi$, i.e., it is not beneficial for the player $P_j$ to construct any edges incident to its associated node $x_j$.
\end{lemma}
\begin{IEEEproof}
Assume by way of contradiction that $|W_j|\ge 1$. Given $G=(V_1 \cup V_2, E_1 \cup E_2 \cup W_j \cup W_{-j})$, we have
\begin{align*}
u_j(W_j \cup W_{-j}|G_1, G_2, G_I) &\le |W_j|(b_j(1)-c_j)+(|I_j|-|W_j|)b_j(2)\\
&= b_j(1)-c_j+(|W_j|-1)(b_j(1)-c_j)+(|I_j|-|W_j|)b_j(2)\\
&\le b_j(1)-c_j+(|I_j|-1)b_j(2)<0,
\end{align*}
where in the above, we are using the fact that $b_j(1)-c_j<b_j(2)$ by the assumption of the lemma. Therefore, we must have that $|W_j|=0$ which yields the required result.
\end{IEEEproof}

In the next result, we propose a condition under which a player disregards the network constructed by another player when considering the best response.  We define the {\it $R$-radius}  of a player $P_i \in P$ with $b_i(1)-c_i>0$ as the minimum integer $R_i>0$ (or $\infty$) such that $b_i(1)-c_i>b_i(R_i+1)$. 

\begin{lemma}\label{lem:surroundedset} 
Consider two players $P_i, P_j \in P$ with {\it R-radii} $R_i$ and $R_j$, respectively. For a given instance of INDG 
$$
\mathcal{G}=(P, (S_i)_{P_i \in P}, (u_i)_{P_i \in P}, G_1, G_2, G_I),
$$
assume that $W_i$ and $W_j$ are  best response actions to $W_{-i}$ and $W_{-j}$, respectively. If $d_{G_1}(x_i, x_j) \ge R_i +R_j -1$, then the actions of the players $P_i$ and $P_j$ are  such that shortest paths from nodes $x_i$ and $x_j$ to the nodes that they depend on in $V_2$ are node disjoint in $G_1$.
\end{lemma}

\begin{IEEEproof}
The idea behind the proof stems from the fact that for any two nodes $x_i,x_j \in V_1$ with $d_{G_1}(x_i,x_j)\ge R_i+R_j-1$, there does not exist any node $x_k \in V_1$ that simultaneously has distance less than $R_i$ to $x_i$ and less than $R_j$ to $x_j$.
To formally prove the lemma, consider $\{(x_i, y_i), (x_j, y_j)\} \subseteq E_I$.  By way of contradiction, assume that the shortest paths from $x_i$ to $y_i$ and $x_j$ to $y_j$ intersect at a node $x_k \in V_1$. Without loss of generality, let $d_{G_1}(x_i, x_k)\ge R_i$. This means that $d_G(x_i, y_i)\ge R_i+1$ where $G=(V_1 \cup V_2, E_1 \cup E_2 \cup W_i \cup W_{-i})$. Now consider $W'_i=W_i \cup \{(x_i,y_i) \}$ as a modified action of player $P_i$. This new action will increase the utility of player $P_i$ by at least $b_i(1)-c_i-b_i(R_i+1)>0$, which is a contradiction to the assumption that $W_i$ is a best response to $W_{-i}$.
\end{IEEEproof}
 
The following example illustrates the application of  Lemma \ref{lem:surroundedset} in determining a Nash equilibrium of the INDG.

\begin{example}
	Consider networks $G_1=(V_1, E_1)$ and $G_2=(V_2, E_2)$ depicted in Fig. \ref{reg1} with the given dependency network $G_I$ between them (shown by dashed edges). Assume that $b_i(3)>b_i(1)-c_i>b_i(4)$ for $i\in \{1,6\}$ which yields $R_1=R_6=3$. Nodes $x_1$ and $x_6$ correspond to the players $P_1$ and $P_6$, respectively. Note that since all of the other nodes $x_i \in V_1\setminus \{x_1, x_2\}$ have $I_i = \emptyset$, their associated players do not construct any edges in any Nash equilibrium by Lemma \ref{lem:number_of_edge}. Both $x_1$ and $x_2$ are dependent on all nodes in $G_2$, as illustrated for $x_1$ in Fig.~\ref{reg2}. 
	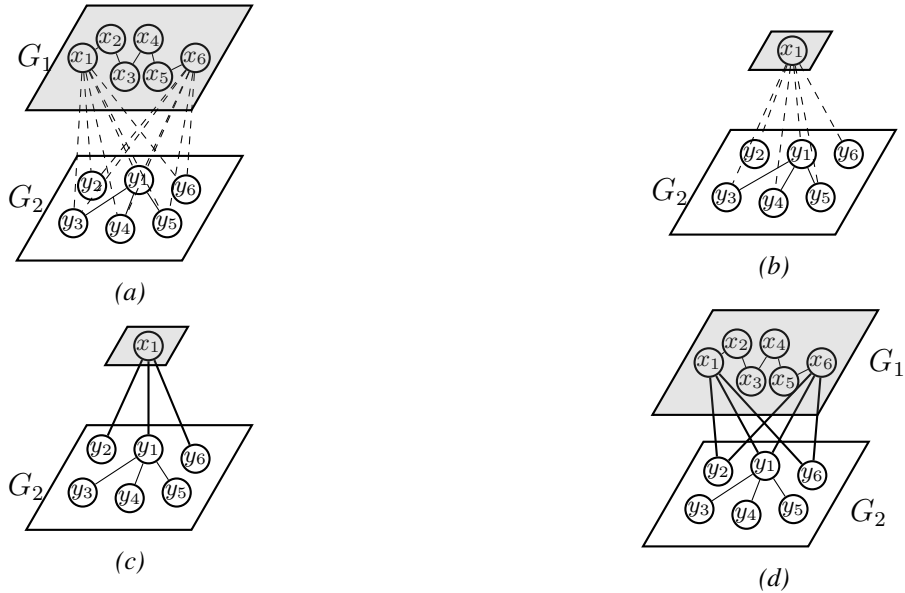
\begin{figure}[h!]
	\begin{minipage}{0.48\textwidth}
		\begin{center}
			\begin{tikzpicture}  [scale=.25, inner sep=0.1pt, minimum size=1pt, auto=center]
			\tikzstyle{man1}=[trapezium, draw, minimum width=3cm, fill=gray,  fill opacity=0.2, trapezium left angle=60, trapezium right angle=120]
			\tikzstyle{man2}=[trapezium, draw, minimum width=3cm,
			trapezium left angle=60, trapezium right angle=120]

			\node[circle, draw=black, thick] (n2) at (-0.5,7)  {\footnotesize $x_1$};

			\node[circle, draw=black, thick] (m1) at (1,8)  {\footnotesize $x_2$};
			\node[circle, draw=black, thick] (m2) at (1.75,6)  {\footnotesize $x_3$};
			\node[circle, draw=black, thick] (m3) at (3,8)  {\footnotesize $x_4$};
			\node[circle, draw=black, thick] (m4) at (3.5,6)  {\footnotesize $x_5$};
			
			\node[circle, draw=black, thick] (n3) at (5.5,7)  {\footnotesize $x_6$};

			\node[circle, draw=black, thick] (t3) at (2.5,0.5) {\footnotesize $y_1$};
			\node[circle, draw=black, thick] (t4) at (0,0.2) {\footnotesize $y_2$};
			\node[circle, draw=black, thick] (t5) at (-1,-1.8)  {\footnotesize $y_3$};
			\node[circle, draw=black, thick] (t6) at (1.5,-2.1) {\footnotesize $y_4$};
			\node[circle, draw=black, thick] (t7) at (4,-1.8)  {\footnotesize $y_5$};
			\node[circle, draw=black, thick][circle, draw=black, thick] (t8) at (5,0)  {\footnotesize $y_6$};

			\draw (n2) -- (m1);
			\draw (m1) -- (m2);
			\draw (m2) -- (m3);
			\draw (m3) -- (m4);
			\draw (m4) -- (n3);
			
			\draw (t3) --(t7);
			\draw (t3) --(t5);
			\draw (t3) --(t6);

			\draw[dashed] (n2) --(t3);
			\draw[dashed] (n2) --(t4);
			\draw[dashed] (n2) --(t5);
			\draw[dashed] (n2) --(t6);
			\draw[dashed] (n2) --(t7);
			\draw[dashed] (n2) --(t8);
			
			\draw[dashed] (n3) --(t3);
			\draw[dashed] (n3) --(t4);
			\draw[dashed] (n3) --(t5);
			\draw[dashed] (n3) --(t6);
			\draw[dashed] (n3) --(t7);
			\draw[dashed] (n3) --(t8);

			\node[circle, draw=black, thick][man1] at (2.5,7) {};
			\node[circle, draw=black, thick][man2] at (2,-1) {};

			\node at (-3,7) {$G_1$};
			\node at (-3.5,-0.5) {$G_2$};

			\end{tikzpicture}    
			\subcaption{}
			\label{reg1}
		\end{center}
	\end{minipage}
	\begin{minipage}{0.48\textwidth}
		\begin{center}
			\begin{tikzpicture}  [scale=.25, inner sep=0.1pt, minimum size=3pt, auto=center]
			\tikzstyle{man1}=[trapezium, draw, minimum width=1.1cm, fill=gray,  fill opacity=0.2, trapezium left angle=60, trapezium right angle=120]
			\tikzstyle{man2}=[trapezium, draw, minimum width=3cm,
			trapezium left angle=60, trapezium right angle=120]

			\node[circle, draw=black, thick] (n2) at (6.5,8)  {\footnotesize $x_1$};
			
			\node[circle, draw=black, thick] (n3) at (7,2.5) {\footnotesize $y_1$};
			\node[circle, draw=black, thick] (n4) at (4.5,2.5) {\footnotesize $y_2$};
			\node[circle, draw=black, thick] (n5) at (3,0.2)  {\footnotesize $y_3$};
			\node[circle, draw=black, thick] (n6) at (5.5,-0.1) {\footnotesize $y_4$};
			\node[circle, draw=black, thick] (n7) at (8,0.2)  {\footnotesize $y_5$};
			\node[circle, draw=black, thick] (n8) at (9.5,2.5)  {\footnotesize $y_6$};

		\draw[dashed] (n2) --(n3);
		\draw[dashed] (n2) --(n4);
		\draw[dashed] (n2) --(n5);
		\draw[dashed] (n2) --(n6);
		\draw[dashed] (n2) --(n7);
		\draw[dashed] (n2) --(n8);

		\draw (n3) --(n7);
		\draw (n3) --(n5);
		\draw (n3) --(n6);
			
			\node[circle, draw=black, thick][man1]  at (6.4,8) {};
			\node[circle, draw=black, thick][man2] at (6,1) {};
			\node at (0,0.5) {$G_2$};

			\end{tikzpicture} 		
			\subcaption{}   
			\label{reg2}
		\end{center}
	\end{minipage}	
	
	\begin{minipage}{0.48\textwidth}
		\begin{center}
			\begin{tikzpicture}  [scale=.25, inner sep=0.1pt, minimum size=3pt, auto=center]
			\tikzstyle{man1}=[trapezium, draw, minimum width=1.1cm, fill=gray,  fill opacity=0.2, trapezium left angle=60, trapezium right angle=120]
			\tikzstyle{man2}=[trapezium, draw, minimum width=3cm,
			trapezium left angle=60, trapezium right angle=120]

			\node[circle, draw=black, thick] (n2) at (6.5,8)  {\footnotesize $x_1$};
			
			\node[circle, draw=black, thick] (n3) at (6.5,2.5) {\footnotesize $y_1$};
			\node[circle, draw=black, thick] (n4) at (4,2.5) {\footnotesize $y_2$};
			\node[circle, draw=black, thick] (n5) at (3,0.2)  {\footnotesize $y_3$};
			\node[circle, draw=black, thick] (n6) at (5.5,-0.1) {\footnotesize $y_4$};
			\node[circle, draw=black, thick] (n7) at (8,0.2)  {\footnotesize $y_5$};
			\node[circle, draw=black, thick] (n8) at (9,2)  {\footnotesize $y_6$};

			\draw[thick] (n2) --(n3);
			\draw[thick] (n2) --(n4);
			\draw[thick] (n2) --(n8);

			\draw (n3) --(n7);
			\draw (n3) --(n5);
			\draw (n3) --(n6);
			
			\node[circle, draw=black, thick][man1]  at (6.4,8) {};
			\node[circle, draw=black, thick][man2] at (6,1) {};
						\node at (0,0.5) {$G_2$};

			\end{tikzpicture} 		
			\subcaption{}   
			\label{reg3}
		\end{center}
	\end{minipage}	
	\begin{minipage}{0.48\textwidth}
		\begin{center}
			\begin{tikzpicture}  [scale=.25, inner sep=0.1pt, minimum size=3pt, auto=center]
			\tikzstyle{man1}=[trapezium, draw, minimum width=3cm, fill=gray,  fill opacity=0.2, trapezium left angle=60, trapezium right angle=120]
			\tikzstyle{man2}=[trapezium, draw, minimum width=3cm,
			trapezium left angle=60, trapezium right angle=120]

			\node[circle, draw=black, thick] (n1) at (3.5,7)  {\footnotesize $x_1$};

			\node[circle, draw=black, thick] (m1) at (5,8)  {\footnotesize $x_2$};
			\node[circle, draw=black, thick] (m2) at (5.75,6)  {\footnotesize $x_3$};
			\node[circle, draw=black, thick] (m3) at (7,8)  {\footnotesize $x_4$};
			\node[circle, draw=black, thick] (m4) at (7.5,6)  {\footnotesize $x_5$};

			\node[circle, draw=black, thick] (n4) at (9.5,7)  {\footnotesize $x_6$};

			\node[circle, draw=black, thick] (t3) at (6.5,1.5) {\footnotesize $y_1$};
			\node[circle, draw=black, thick] (t4) at (4,1.2) {\footnotesize $y_2$};
			\node[circle, draw=black, thick] (t5) at (3,-0.8)  {\footnotesize $y_3$};
			\node[circle, draw=black, thick] (t6) at (5.5,-1.1) {\footnotesize $y_4$};
			\node[circle, draw=black, thick] (t7) at (8,-0.8)  {\footnotesize $y_5$};
			\node[circle, draw=black, thick][circle, draw=black, thick] (t8) at (9,1)  {\footnotesize $y_6$};

			\draw (n1) -- (m1);
			\draw (m1) -- (m2);
			\draw (m2) -- (m3);
			\draw (m3) -- (m4);
			\draw (m4) -- (n4);
			
			\draw (t3) --(t7);
			\draw (t3) --(t5);
			\draw (t3) --(t6);
			
			\draw[thick] (n1)--(t3);
			\draw[thick] (n1)--(t4);
			\draw[thick] (n1)--(t8);

			\draw[thick] (n4) --(t3);
			\draw[thick] (n4) --(t4);
			\draw[thick] (n4) --(t8);

			\node[circle, draw=black, thick][man1] at (6.5,7) {};
			\node[circle, draw=black, thick][man2] at (6,0) {};

			\node at (13,7) {$G_1$};
			\node at (12,-1) {$G_2$};

			\end{tikzpicture}    
			\subcaption{}
			\label{reg4}
		\end{center}
	\end{minipage}
				
	\caption{(a) Networks $G_1$ and $G_2$ with interdependencies shown by dashed edges. (b) Interdependencies of player $P_1$ with nodes in $G_2$. (c) Best response action of $P_1$ (d) A Nash equilibrium instance.}
	\label{reg}
\end{figure}
	The distance between nodes $x_1$ and $x_6$ in $G_1$ is $5$ and thus the networks constructed by players $P_1$ and $P_6$ will be such that the shortest paths from $x_1$ to the nodes in $G_2$ are node disjoint (in $G_1$) from the shortest paths from $x_6$ to the nodes in $G_2$, by Lemma~\ref{lem:surroundedset}. Fig.~\ref{reg3} demonstrates a best response for player $P_1$. Using the optimal action of $P_1$ and Lemma~\ref{lem:surroundedset}, we can determine a Nash equilibrium as shown in Fig.~\ref{reg4}.
\end{example}

\section{Nash Equilibrium of INDG for Networks Containing Star Subgraphs}
\label{sec:dist_nash}

With our results on best responses in hand, we now turn our attention to proving the existence of a Nash equilibrium.  While it is challenging to show this for general $G_1, G_2$ and $G_I$, we will prove that the INDG always has a Nash equilibrium when $G_2$ contains a star subgraph,\footnote{Such networks can be used to represent, for example, sensor networks that have a fusion center, or transportation networks that have a ``hub-and-spoke'' structure \cite{sensorwireless,USPS,jeanpaul, ATN}.} and $G_I=(V_1 \cup V_2, E_I)$ is the complete bipartite network, i.e., $E_I=V_1 \times V_2$.  We allow $G_1$ to be arbitrary.  Without loss of generality, let $y_1 \in V_2$ be a hub node in $G_2=(V_2, E_2)$, i.e., $(y_1, y) \in E_2 ~ \forall y \in V_2 \setminus \{y_1\}$.
As we illustrate later, the presence of heterogeneous players (captured by  individual benefit functions and edge costs) along with the arbitrary structure of $G_1$ leads to non-trivial interconnection networks in equilibrium, even under the above  assumptions on $G_2$ and $G_I$.

To develop our results, we partition the set of players $P$ into two sets: {\it high-cost players}
\begin{equation}\label{def:highcost}
S_H = \{P_i \in P |b_i(1) -b_i(2) < c_i\},
\end{equation}
and {\it low-cost players}
\begin{equation}\label{def:lowcost}
S_L = \{P_i \in P |b_i(1)-b_i(2) \ge c_i\}.
\end{equation}
Recall that we assumed $V_1=\{x_{1}, \cdots, x_{n}\}$ and $V_{2}=\{y_{1}, \cdots, y_{m}\}$. For the rest of this section, we denote the number of players $|P|$ by $|V_1|=n$ and the number of nodes in $|V_2|$ by $m$.

We begin our analysis in this section with the following useful corollary of Lemma~\ref{lem:lowlowcost}, which determines a best response action for the low-cost players. 

\begin{corollary}\label{cor:lowcost}
Assume that $P_i \in S_L$. Then $W_i=\{(x_i, y)|y \in V_2\}$ is a best response action for player $P_i$ regardless of the actions of the other players.
\end{corollary}

For the rest of this section, we assume that low-cost players always set their action according to the best response given by Corollary~\ref{cor:lowcost}.
In the next proposition, we discuss the best responses of high-cost players when there is a low-cost player in their neighborhood. We define the {\it L-radius} of a player $P_i \in S_H$ as the maximum nonnegative integer $L_i$ such that\footnote{A player $P_i$ with $b_i(1)-c_i+(m-1)b_i(2)<0$ is defined to have $L_i=\infty$, and his/her best response action is always the empty network by Lemma~\ref{lem:threshold}.} 
\begin{equation}\label{L_value}
b_i(1)-c_i+(m-1)b_i(2) \le mb_i(L_i+1).
\end{equation}

\begin{proposition}\label{prop:hig_cost_vicinity}
Let $P_i$ be a high-cost player. Suppose that there exists a low-cost player $P_j \in S_L$ such that the distance between $x_j$ and $x_i$ is less than $L_i+1$ (i.e., $d_{G_1}(x_i, x_j)<L_i+1$), where $L_i$ is the $L$-radius of player $P_i$. Then, if $P_j$ has constructed edges to all nodes in $V_2$, the empty network is a best response action for player $P_i$.
\end{proposition}

\begin{IEEEproof}
Let $W_i$ denote a best response action of the player $P_i \in S_H$ with respect to $W_{-i}$. Node $x_i$ has distance $d \le L_i$ to $x_j$ which is associated with a low-cost player $P_j \in S_L$ that is connected to all of the nodes in $V_2$.  
Now assume that $W_i \neq \phi$. Then we have
\begin{align} \nonumber
\Psi_i(W_1, \cdots, W_{n})&= u_i(\cup_{j =1}^{n}W_j|G_1, G_2, G_I)\\ \nonumber
&= \left( \sum_{y_j \in I_i} b_i(d_{G}(x_i,y_j))\right) -c_i|W_i|\\ \label{hig_cost_vicinity}
&\le |W_i|(b_i(1)-c_i)+(m-|W_i|)b_i(2)\\\nonumber
&=b_i(1)-c_i+(|W_i|-1)(b_i(1)-c_i)+(m-|W_i|)b_i(2)\\\nonumber
&\le b_i(1)-c_i+(|W_i|-1)b_i(2)+(m-|W_i|)b_i(2)\\\nonumber
&= b_i(1)-c_i+(m-1)b_i(2)\\\nonumber
&\le mb_i(L_i+1) \le mb_i(d+1).
\end{align}
Therefore, player $P_i$ can increase its utility by changing its action to be the empty network and connecting to the nodes it depends on in $G_2$ via edges constructed by the low-cost player $P_j$. 
\end{IEEEproof}

The above result shows that the existence of a low-cost player in the proximity of a high-cost player  will make the high-cost player a {\it free rider} in any Nash equilibrium, i.e., the high-cost player does not construct any edges, and instead benefits from the low-cost player's edges. 

\begin{remark}
	Note that Corollary~\ref{cor:lowcost} and Proposition~\ref{prop:hig_cost_vicinity} do not rely on $G_2$ having a star subgraph, and hold whenever the low-cost players have dependencies on all nodes in $G_2$.
\end{remark}

\begin{corollary}\label{cor:highcost_action}
Assume that $P_i \in S_H$. Then for any best response action of the player $P_i$, node $x_i$ is either connected to only the center of a star subgraph in $G_2$ (e.g., node $y_1 \in V_2$) or it does not have any edges. 
\end{corollary}

\begin{IEEEproof}
Since $G_2$ has a star subgraph, the size of its smallest dominating set is 1 (e.g., the center of the star, $y_1$). Therefore, by Lemma \ref{lem:highcost_action_changed}, we must have that $|W_i| \le 1$. Furthermore, the proof of Lemma \ref{lem:highcost_action_changed} shows that with a single edge, $W_i=\{(x_i,y_1)\}$ produces the highest possible utility for $P_i$. Proposition \ref{prop:hig_cost_vicinity} gives an instance of the situation when $W_i=\phi$. 
\end{IEEEproof}

Although Corollary \ref{cor:highcost_action} limits the set of best response actions of a high-cost player to two actions (namely, connect to a hub in $G_1$ or not), it is not clear whether this game has a pure strategy Nash equilibrium for any set of players with arbitrary network $G_1$, edge cost $c_i$ and benefit function $b_i(\cdot)$. We prove existence of a pure Nash equilibrium in this game by providing an algorithm that outputs such an equilibrium. To do this, we first define an index $r_i$ for each high-cost player $P_i \in S_H$, called the {\it $r$-radius}. The $r$-radius of player $P_i$ with benefit function $b_i(\cdot)$ and edge cost $c_i$ is defined as the maximum nonnegative integer $r_i$ such that\footnote{A player $P_i$ with $b_i(1)-c_i+(m-1)b_i(2)<0$ is defined to have $r_i=\infty$, and his/her best response action is always the empty network by Lemma~\ref{lem:threshold}.}
\begin{align}\label{radius}
b_i(1)-c_i+&(m-1)b_i(2) \le b_i(r_i+1)+(m-1)b_i(r_i+2).
\end{align}
Note that by the above definition, $L_i \ge r_i$ where $L_i$ was defined in \eqref{L_value}.
For a given $r$-radius $r_i$, we define the $r_i$-neighborhood of node $x_i$ as 
\begin{align}\label{N_i}
N_i=\{x_j|P_j \in S_H ~\text{and}~ d_{G_1}(x_i,x_j) \le r_i\}.
\end{align}
If a high-cost player $P_i$ has another high-cost player $P_j$ with a single edge to a hub node in $V_2$ such that $x_j \in N_i$, then player $P_i$ is better off with no edge to $V_2$. This statement is also true if $P_j$ is a low-cost player by Proposition~\ref{prop:hig_cost_vicinity} and the fact that $r_i \le L_i$.  The following proposition formally states these ideas.

\begin{proposition}\label{prop:high_highcost_vicinity}
Let $P_i$ be a high-cost player with $r$-radius $r_i$. Suppose that there exists a player $P_j$ such that $x_j$ is connected to a hub node in $V_2$ and $d_{G_1}(x_i,x_j) \le r_i$.  Then the empty network is a best response action for the player $P_i$ with respect to $W_{-i}$.
\end{proposition}

The results that we provided in this section enable us to give an algorithm that outputs a Nash equilibrium instance of the interconnection network design game with distance utilities for an arbitrary network $G_1$ and arbitrary benefit function and cost of edges.

\begin{theorem}\label{thm:nash}
Assume that network $G_2$ has a star subgraph and $G_I$ is a complete bipartite graph with partitions $V_1$ and $V_2$. Then the interconnection network design game with distance utilities in Definition \ref{def:INDG_DU} always has  a pure strategy Nash equilibrium. 
\end{theorem}

\begin{IEEEproof}
We prove this theorem by providing an algorithm that outputs a Nash equilibrium instance of the game given by a set of actions $(W_1, W_2, \cdots, W_n)$ for the players. The steps of the algorithm are as follows:
\begin{enumerate}
\item  
Connect nodes associated to the low-cost players to all of the nodes in $V_2$.
\item Take $S_H^{\infty}$ as the set of all high-cost players with $r_i=\infty$ (which includes all of the players with $L_i=\infty$). Set the actions of all players $P_i \in S_H^{\infty}$ to be the empty network, i.e., $W_i = \emptyset$. 
\label{alg:r_infinite}
\item Determine the set $S_H^L$ which consists of all high-cost players that have a low-cost player in their $L_i$-neighborhood where $L_i$ denotes the $L$-radius, i.e., 
$$
S_H^L=\{P_i \in S_H|\exists P_j \in S_L \hbox{ such that } d_{G_1}(x_i,x_j) \le L_i\}.
$$
Set the actions of these players to be the empty network (by Proposition \ref{prop:hig_cost_vicinity}).
\item Let $Q \subseteq S_H \setminus (S^L_H \cup S_H^{\infty})$ be the set of players whose actions have not been determined yet. If the set $Q$ is empty, exit the algorithm. Otherwise, let $P_i \in Q$ be the player with the lowest $r$-radius. Connect $x_i$ (i.e., the node associated to $P_i$) via a single edge to a central node in $G_2$. Remove $P_i$ from $Q$. \label{alg:sort-step}
\item Set the action of all high-cost players $P_j \in Q$ with $x_i \in N_j$ to the empty network and remove them from the set $Q$. Recall that $N_j$ is the $r_j$-neighborhood of player $P_j$ and was defined in \eqref{N_i}.\label{middle-step}
\item Return to step \ref{alg:sort-step}. \label{alg:last-step}
\end{enumerate}
We now argue that the output of the above algorithm is in fact a Nash equilibrium.  Since the actions of low-cost players are in accordance with Corollary~\ref{cor:lowcost}, they are best responses to the other actions. Next, note that if a high cost player  $P_i \in S_H^{\infty}$ wants to take a best response action with respect to the actions of the other players, he/she has to choose the empty network by Lemma \ref{lem:threshold}; this is the action that our proposed algorithm assigns to these players. The same is true for all high cost players with a low-cost player in their $L_i$-neighborhood, according to Proposition~\ref{prop:hig_cost_vicinity}. Thus we only need to prove optimality of the actions of the remaining players which are determined through steps \ref{alg:sort-step} to \ref{alg:last-step}. Note that all of the remaining players have $r_i < \infty$.

Consider the set $S_H \setminus (S^L_H \cup S_H^{\infty})=\{P_{i_1}, \cdots, P_{i_t}\}$ and assume without loss of generality that $r_{i_1} \le r_{i_2} \le \cdots \le r_{i_t}$. Under the algorithm, the action of $P_{i_1}$ is $W_{i_1}=\{(x_{i_1},y_1)\}$. We know that there is no low-cost player in the $L_{i_1}$-neighborhood of $P_{i_1}$, since $P_{i_1} \in S_H \setminus (S^L_H \cup S_H^{\infty})$. Similarly, there is no high-cost player in $S_H^L \cup S_H^{\infty}$ with a nonempty action in the $N_{i_1}$ neighborhood of $P_{i_1}$. Now assume that there exists a player $P_{i_j} \in S_H \setminus (S^L_H \cup S_H^{\infty})$ with $j>1$ and $|W_{i_j}|=1$. We have to show that $x_{i_j} \notin N_{i_1}$, since otherwise the action of player $P_{i_1}$ will not be optimal. In step \ref{middle-step} of the algorithm, we set the actions of all players $P_{i_q}$ such that $x_{i_1} \in N_{i_q}$ to the empty network and remove them from the set $Q$. Hence, we must have that $x_{i_1} \notin N_{i_j}$, i.e., $d_{G_1}(x_{i_1}, x_{i_j})>r_{i_j} \ge r_{i_1}$. Therefore, $x_{i_j} \notin N_{i_1}$ and thus the action of $P_{i_1}$ is optimal.

The actions of all players that construct the empty network in step \ref{middle-step} are optimal, by Proposition~\ref{prop:high_highcost_vicinity}.

Finally, consider any player $P_{i_j}$ with $|W_{i_j}|=1$ and $j>1$. We know that $x_{i_k} \notin N_{i_j}$ for any $k<j$ with $|W_{i_k}|=1$; otherwise the action of player $P_{i_j}$ would have been set to the empty network in step \ref{middle-step} of the algorithm after assigning the action of player $P_{i_k}$ in step \ref{alg:sort-step}. Moreover, by a reasoning similar to the argument for optimality of $P_{i_1}$'s action, we can show that for any player $P_{i_t}$ with $t>j$ and  $|W_{i_t}|=1$,  we have $x_{i_t} \notin N_{i_j}$. Therefore, the action of player $P_{i_j}$ is a best response.

Thus, each player is playing their best response given the actions of the rest of the players, which implies that the given vector of actions is a Nash equilibrium.  
\end{IEEEproof}

\begin{remark}
Note that the algorithm provided in the proof of the above theorem corresponds to {\it sequential best response dynamics} by the players, where they move in the order specified by the algorithm.
\end{remark}

The following example illustrates the steps of the algorithm, and the corresponding Nash equilibrium.

\begin{figure}[h!]
	\begin{minipage}{0.48\textwidth}
		\begin{center}
			\begin{tikzpicture}  [scale=.35, inner sep=0.1pt, minimum size=3pt, auto=center]
			
			\node[circle, draw=black, thick] (n1) at (1,1) {\footnotesize $x_1$};
			\node[circle, draw=black, thick] (n2) at (5,-1) {\footnotesize $x_2$};
			\node[circle, draw=black, thick] (n3) at (6,3)  {\footnotesize $x_3$};
			\node[circle, draw=black, thick] (n4) at (7,6) {\footnotesize $x_4$};
			\node[circle, draw=black, thick] (n5) at (8,4)  {\footnotesize $x_5$};
			\node[circle, draw=black, thick] (n6) at (4,5)  {\footnotesize $x_6$};
\node[circle, draw=black, thick] (n7) at (3,0)  {\footnotesize $x_7$};
\node[circle, draw=black, thick] (n8) at (3,3)  {\footnotesize $x_8$};
			\node[circle, draw=black, thick] (n9) at (7,0)  {\footnotesize $x_9$};
			\draw (n8) --(n1);
			\draw (n8) --(n7);
			\draw (n7) --(n3);
			\draw (n3) --(n5);
			\draw (n3) --(n6);
			\draw (n3) --(n2);
			\draw (n6) --(n4);
			\draw (n2) --(n9);
			\node at (-1,3) {$G_1$};

			\end{tikzpicture}    
			\subcaption{}
			\label{fig_1}
		\end{center}
	\end{minipage}
	\begin{minipage}{0.48\textwidth}
		\begin{center}
			\begin{tikzpicture}  [scale=.35, inner sep=0.1pt, minimum size=3pt, auto=center]
		
			\node[circle, draw=black, thick] (n11) at (7.5,0) {\footnotesize $y_1$};
			\node[circle, draw=black, thick] (n12) at (5,1.5) {\footnotesize $y_2$};
			\node[circle, draw=black, thick] (n13) at (4,0.2)  {\footnotesize $y_3$};
			\node[circle, draw=black, thick] (n14) at (2.7,-1) {\footnotesize $y_4$};
			\node[circle, draw=black, thick] (n15) at (1.5,-2.5)  {\footnotesize $y_5$};
			\node[circle, draw=black, thick] (n16) at (0.5,-4.2)  {\footnotesize $y_6$};
			\node[circle, draw=black, thick] (n17) at (0,-6)  {\footnotesize $y_7$};
			
			\draw (n11) --(n12);
			\draw (n11) --(n13);
			\draw (n11) --(n14);
			\draw (n11) --(n15);
			\draw (n11) --(n16);
			\draw (n11) --(n17);
			\node at (8.5,-2) {$G_2$};

			\end{tikzpicture}    
			\subcaption{}
			\label{fig_2}
		\end{center}
	\end{minipage}
\caption{(a) Network $G_1$ (b) Network $G_2$. }
\end{figure}
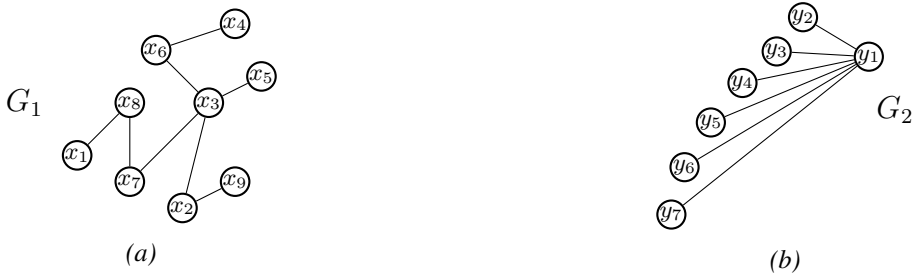

\begin{example}\label{examp1}
Consider two networks $G_1=(V_1, E_1)$ and $G_2=(V_2, E_2)$ depicted in Figures \ref{fig_1} and \ref{fig_2} with complete dependencies between nodes in $G_1$ and $G_2$. Assume that the cost of constructing edges is equal to 1 for all of the players, i.e., $c_i=1, 1 \le i \le 9$. Suppose the benefit functions for the players take the values given in Table~\ref{tab:example1}. Based on these values, player 7 is a low-cost player (since $c_7 < b_7(1)-b_7(2)$) and the rest of the players have high edge costs, i.e.,
\begin{align*}
&S_L=\{P_7\},\\
&S_H=\{P_1, P_2, \cdots, P_6, P_8, P_9\}.
\end{align*}
The corresponding values of the radii $r_i$ and $L_i$ (given by inequalities \eqref{radius} and \eqref{L_value}, respectively) are shown in the table. We now follow the algorithm prescribed in the proof of Theorem~\ref{thm:nash}.

\begin{table}
	\begin{center}
		\begin{tabular}{| l | l | l | l  | l | l | l | l|}
			\hline
			~~~   & $b_i(1)$ & $b_i(2)$ & $b_i(3)$ & $b_i(4)$ & $b_i(5)$ & $L_i$ & $r_i$\\ \hline
			$P_1$ & 1.5 & 1.3 & 1.2 & 1.1 & 0.2 & 2 & 1 \\ \hline
			$P_2$ & 1.2 & 0.8 & 0.5 & 0.2 & 0 & 1 & 0  \\ \hline
			$P_3$ & 1.1 & 0.9 & 0.1 & 0 & 0 & 1 & 0 \\ \hline
			$P_4$ & 0.9 & 0.8 & 0.7 & 0.5 & 0.2 & 2 & 1  \\ \hline
			$P_5$ &  1.2 & 1.1 & 0.9 & 0.2 & 0.1 & 1 & 0   \\ \hline
			$P_6$ &  1.3 & 1 & 0.5 & 0.4 & 0.3 & 1 & 0 \\ \hline
			$P_7$ & 3 & 1 & 0.5 & 0.5 & 0.4 & NA & NA\\ \hline
			$P_8$ &  1.2 & 0.8 & 0.7 & 0.5 & 0.4 & 1 & 1  \\ \hline
			$P_9$ &  1.2 & 1.1 & 1.1 & 1 & 0.2 & 3 & 2 \\ \hline
		\end{tabular}
	\end{center}
	\caption{Benefit function, $r$-radius and $L$-radius of the players in Example \ref{examp1}.}
	\label{tab:example1}
\end{table}
\begin{enumerate}
	\item $P_7$ is the only low-cost player, and thus we connect $x_7$ to all of the nodes in $G_2$, i.e., $W_7=\{(x_7, y_i)| 1 \le i \le 7\}$.
	\item For each node $v_i$ whose distance to the low-cost player $x_7$ is at most $L_i$, we set that player's action to be empty.  These nodes are given by $\{P_1, P_3, P_8, P_9\}$, and thus  $W_1=W_3=W_8=W_9=\emptyset$.
	\item The second player has the lowest $r$-radius among the remaining players and thus we set its action to $W_2=\{(x_2, y_1)\}$. Since $\nexists P_j, j \in \{4,5,6\}$ such that $x_2 \in N_j$, we must choose the next player with the lowest $r_i$. Recall that $N_j$ was defined in \eqref{N_i}.
	\item Player $P_5$ with $r_5=0$ has the lowest $r$-radius among the remaining players. Thus we set $W_5=\{(x_5, y_1)\}$. Again since $\nexists P_j, j \in \{4,6\}$ such that $x_5 \in N_j$, we must choose the next player with the lowest $r_i$.
	\item Finally, we choose player $P_6$ with $r_6=0$ and set its action to  $W_6=\{(x_6, y_1)\}$. Due to the fact that $x_6 \in N_4$, we set the action of player $P_4$ to the empty network, i.e., $W_4=\phi$.
\end{enumerate}
Fig. \ref{fig3} demonstrates the output of the algorithm given in the proof of Theorem \ref{thm:nash} when networks $G_1$ and $G_2$ depicted in Fig. \ref{fig_2} are given as input.
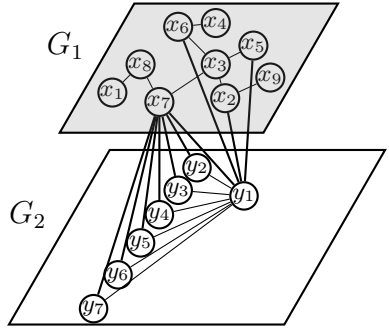
\begin{figure}[h!]

	\begin{center}
		\begin{tikzpicture}  [scale=.25, inner sep=0.01pt, minimum size=3pt, auto=center]
		\tikzstyle{man1}=[trapezium, draw, minimum width=3.7cm, fill=gray,  fill opacity=0.2, trapezium left angle=60, trapezium right angle=120]
		\tikzstyle{man2}=[trapezium, draw, minimum width=5cm,
		trapezium left angle=60, trapezium right angle=120]
		
\node[circle, draw=black, thick] (n1) at (0.5,0.5) {\footnotesize $x_1$};
\node[circle, draw=black, thick] (n2) at (6.5,0.2) {\footnotesize $x_2$};
\node[circle, draw=black, thick] (n3) at (6,2)  {\footnotesize $x_3$};
\node[circle, draw=black, thick] (n4) at (6,4.2) {\footnotesize $x_4$};
\node[circle, draw=black, thick] (n5) at (8,3)  {\footnotesize $x_5$};
\node[circle, draw=black, thick] (n6) at (4,4)  {\footnotesize $x_6$};
\node[circle, draw=black, thick] (n7) at (3,0)  {\footnotesize $x_7$};
\node[circle, draw=black, thick] (n8) at (2,2)  {\footnotesize $x_8$};
\node[circle, draw=black, thick] (n9) at (8.9,1.3)  {\footnotesize $x_9$};
\draw (n8) --(n1);
\draw (n8) --(n7);
\draw (n7) --(n3);
\draw (n3) --(n5);
\draw (n3) --(n6);
\draw (n3) --(n2);
\draw (n6) --(n4);
\draw (n2) --(n9);

			\node[circle, draw=black, thick] (n11) at (7.5,-5) {\footnotesize $y_1$};
			\node[circle, draw=black, thick] (n12) at (5,-3.5) {\footnotesize $y_2$};
			\node[circle, draw=black, thick] (n13) at (4,-4.7)  {\footnotesize $y_3$};
			\node[circle, draw=black, thick] (n14) at (3,-6) {\footnotesize $y_4$};
			\node[circle, draw=black, thick] (n15) at (2,-7.5)  {\footnotesize $y_5$};
			\node[circle, draw=black, thick] (n16) at (0.8,-9.2)  {\footnotesize $y_6$};
			\node[circle, draw=black, thick] (n17) at (-0.5,-11)  {\footnotesize $y_7$};
			
			\draw (n11) --(n12);
			\draw (n11) --(n13);
			\draw (n11) --(n14);
			\draw (n11) --(n15);
			\draw (n11) --(n16);
			\draw (n11) --(n17);

			\draw[thick] (n7) --(n11);
			\draw[thick] (n7) --(n12);
			\draw[thick] (n7) --(n13);
			\draw[thick] (n7) --(n14);
			\draw[thick] (n7) --(n15);
			\draw[thick] (n7) --(n16);
			\draw[thick] (n7) --(n17);
			
			\draw[thick] (n6) --(n11);
			\draw[thick] (n2) --(n11);
			\draw[thick] (n5) --(n11);

		\node[circle, draw=black, thick][man1] at (5.1,1.8) {};
		\node[circle, draw=black, thick][man2] at (5,-7.2) {};

		\node at (-2,3) {$G_1$};
		\node at (-4,-6) {$G_2$};

		\end{tikzpicture}    
	\end{center}

	\caption{Networks $G_1$ and $G_2$ with the Nash equilibrium interconnection network $G_p$  connecting them. Network $G_p$ was produced by the algorithm given in the proof of Theorem \ref{thm:nash}.}
	\label{fig3}
\end{figure}
\end{example}

As one can see, the role of the hub nodes is crucial in the structure of the Nash equilibrium interconnection networks. While low edge cost players connect their associated nodes in $G_1$ to all of the nodes in the network $G_2$ (and thus themselves become hubs), the remaining high-cost players either choose (I) the empty network and connect via edges constructed by other players or (II) they directly connect to the hub node in network $G_2$. 

\subsection{Price of Anarchy}
The concept of ``price of anarchy'' (PoA) was introduced in \cite{papad} to measure how selfish behavior of the individual players  degrades the efficiency of the output in a non-cooperative game. Given a strategy $W=(W_1, W_2, \ldots, W_n)$ taken by the players and $T(W)=\sum_{i=1}^n u_i(W_i \cup W_{-i}|G_1, G_2, G_I)$ as the social welfare function, PoA is defined as
$$
PoA=\frac{\max_{W \in S} T(W)}{\min_{W \in E} T(W)},
$$
where  $S$ denotes the joint strategy space and $E \subseteq S$ is the set of strategies in Nash equilibrium.

We show via the following example that the PoA can be arbitrarily large in the INDG.

\begin{example}
Consider two networks $G_1=(V_1, E_1)$ and $G_2=(V_2, E_2)$, each containing  star subgraphs centered on nodes $x_1 \in V_1$ and $y_1 \in V_2$, respectively, i.e., 
\begin{align*}
\{(x_1, x_i)|x_i \in V_1\setminus\{x_1\}\} \subseteq E_1\\
\{(y_1, y_i)|y_i \in V_2\setminus\{x_2\}\} \subseteq E_2.
\end{align*}
Suppose that we have full dependencies between nodes in $V_1$ and $V_2$, i.e., $G_I=(V_1 \cup V_2, V_1 \times V_2)$. Assume that all of the players $P_i, 1 \le i \le |V_1|$ in the INDG have $c_i=2.1, b_i(1)=1$ and $b_i(2)=b_i(3)=1/(|V_2|-1)$, where $|V_1|>|V_2| >1$. This means that $b_i(1)-c_i+(|V_2|-1)b_i(2)=-0.1<0$ for all of the players $P_i \in P$ and thus by Lemma \ref{lem:threshold}, none of the players constructs any edges. Therefore, the social welfare value is $T(W)=0$ for all strategies in Nash equilibrium.

Now consider the socially optimal interconnection strategy, i.e., the strategy that maximizes $T(\cdot)$. For the strategy $W^\star=(W^\star_1, W^\star_2,\ldots, W^\star_n)$  where $W^\star_1=\{(x_1,y_1)\}$ and $W^\star_i=\phi, i\neq 1$, we have that 
\begin{align*}
T(W^\star)&=\sum_{i=1}^n u_i(W^*_i \cup W^*_{-i}|G_1, G_2, G_I)\\\
&=b_1(1)-c_1+(|V_2|-1)b_1(2)+\sum_{j=2}^{|V_1|}\left(b_j(2)+(|V_2|-1)b_j(3)\right)\\
&=-0.1+\frac{(|V_1|-1)|V_2|}{|V_2|-1}>0.
\end{align*}
Therefore, the network that maximizes the social welfare function has a nonzero utility and thus PoA is trivially infinite.
\end{example}

\begin{remark}
The network $G_{Soc Opt}=\cup_{i=1}^n W_i$ that maximizes the social utility function $T(W)$ is called the socially optimal network. Similar to the proof of the Theorem \ref{thm:dis_nphard}, one can show that finding the socially optimal network is an NP-hard problem. 
\end{remark}

\begin{remark}
When all players have low costs for constructing edges, one can show that the Nash equilibrium networks are also socially optimal.
\end{remark}

\section{Simulation and Numerical Analysis}
In this section, we use our algorithm from the previous section to investigate the Nash equilibria that arise in large interconnected cyber-physical systems consisting of a power network and a sensor/communication network (as described in Section~\ref{sec:intro}). Here, we use the same experimental setup as \cite{cyberpower}, where networks $G_1$ and $G_2$ have a synthetic scale-free (SF) structure. Such scale-free networks have  also been used to model power and communication networks in many of the other works in this area \cite{porter,P17,buldyrev2010}.

In order to provide comparisons and insights, we also consider the cases that the power network is an Erdos-Renyi (ER) random network, or a geometric random network. In ER random networks, each edge is placed independently with a fixed probability \cite{P1}. Although ER networks are not typically representative of real-world networks, they are a common baseline model for studying large scale networks   \cite{P4, P1}. Geometric random (GR) networks consist of a set of spatially distributed nodes, and there is an edge between two nodes if their distance (in some metric) is less than a given threshold \cite{geometric}.

\begin{figure}
    \centering
    \begin{subfigure}[b]{0.3\textwidth}
        \includegraphics[width=\textwidth]{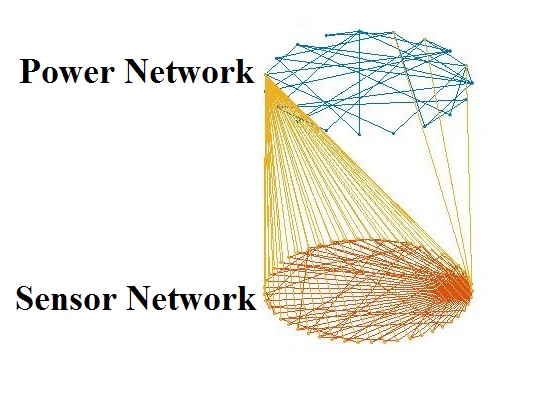}
        \caption{Scale-Free Network}
        \label{fig:gull}
    \end{subfigure}
    ~ 
    \begin{subfigure}[b]{0.3\textwidth}
        \includegraphics[width=\textwidth]{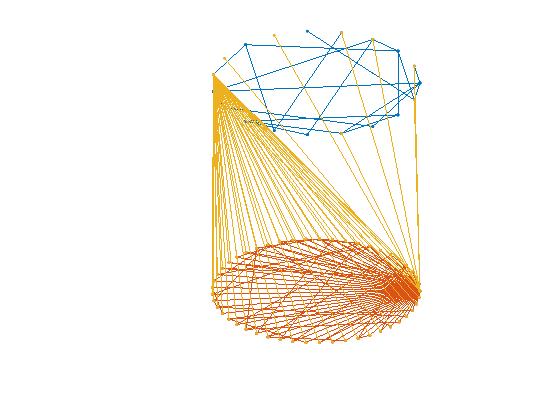}
        \caption{Erdos-Renyi Network}
        \label{fig:tiger}
    \end{subfigure}
    ~ 
    \begin{subfigure}[b]{0.3\textwidth}
        \includegraphics[width=\textwidth]{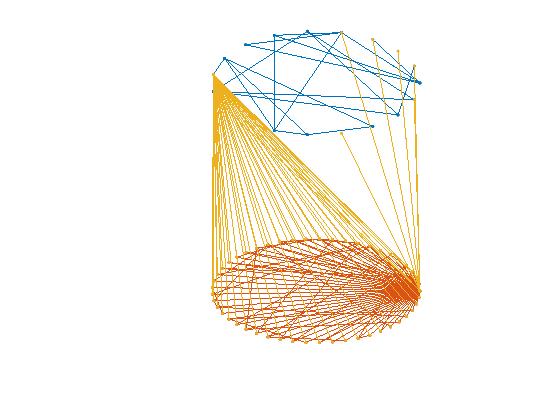}
        \caption{Geometric Random Network}
        \label{fig:mouse}
    \end{subfigure}
    \caption{Output of the proposed algorithm in Theorem \ref{thm:nash} when the communication network has SF structure and the power network has (a) SF, (b) ER and (c) GR structure. Note that in order to keep the figures clean and the structures visible, we have decreased the number of nodes in the power and communication networks to 20 and 50 nodes, respectively.}\label{sim_ind}
\end{figure}

We consider the case that there are 500 power substations that supply electricity to 5000 regions. Therefore, there are 500 and 5000 nodes in the power ($G_1$) and the sensor ($G_2$) networks, respectively. We create a hub node in the sensor network by connecting an arbitrary node in $V_2$ to all of the other nodes in $V_2$. Furthermore, we assume that the interdependency network $G_I$ is a full bipartite network, i.e., $E_I=V_1 \times V_2$.
We consider the same benefit function for all of the players in the game, i.e., $b_i(\cdot)=b(\cdot)$ for all $1 \le i \le 500$. In our simulation, we set $b(1)=1.2, b(2)=0.7, b(3)=0.6, b(4)=0.5, b(5)=0.3, b(6)=0.2$ and $b(k)=0$ for $k \ge 7$. In order to model the situation that we have a wide range of players, we choose the cost of constructing edges for the players uniformly at random between 0.01 and 2500. We set the parameters of the SF, ER and GR networks such that they all have approximately the same number of edges. We consider three scenarios:
\begin{enumerate}
	\item The power network is a SF  network constructed by preferential attachment \cite{barabasi}, with 5 initial nodes. Each newly added node connects to six existing nodes.
	\item The power network is an ER network with edge formation probability of 0.024. 
	\item The power network is a GR network in which nodes are uniformly distributed in a $2 \times 2$ square, and the threshold to form an edge is set at a distance of  0.18.
\end{enumerate}
In all of the above cases, we assume that the sensor network has a fixed SF structure constructed by preferential attachment with 5 initial nodes, and each newly added node connects to one of the existing nodes. Figure \ref{sim_ind} demonstrates the power and sensor networks and the set of Nash equilibrium interconnection edges between them (we have reduced the number of nodes to keep the structures visible). These edges are constructed according to the algorithm in the proof of Theorem \ref{thm:nash}. 
In Figure \ref{sim_ind}, there is only one low cost player; by Corollary \ref{cor:lowcost}, we know that such players always construct edges to all of the nodes in $G_2$ and this is independent of the actions of the other players or structures of the networks $G_1$ and $G_2$. We have summarized some of the salient features of the Nash equilibrium interconnected networks in Table \ref{tab:sim}. These results are produced by averaging over 100 instances of the random networks and player edge costs.  As the data in the table indicates, the number of constructed interconnection edges (and the corresponding social welfare and distances between nodes) is approximately the same in all of the three forms of the power networks, namely SF, ER and GR.   This suggests that for the setting considered here (involving random power networks and uniformly distributed edge costs across the nodes), the topology of the power network does not significantly impact the characteristics of the Nash equilibrium interconnection network.  

\begin{table*}[t]
	\begin{center}
		\begin{tabular}{| l | l | l | l|}
			\hline
			~~~   & SF & ER & GR\\ \hline
			$|E_1|$ & 2970.7 & 2980.8 & 2945.1 \\ \hline
			Diameter of $G_1$ & 4 & 4.42 & $\infty$ \\ \hline
			Total Number of Interconnection Edges Constructed & 5106.7 & 5103.5 & 5104.7  \\ \hline
			Interconnection Edges Constructed by High Cost Players & 106.7 & 103.5 & 104.7 \\ \hline
			Average Distance Between Interdependent Nodes & 4.00 & 4.00 & 4.00 \\ \hline
			Social Welfare of the Interconnected Network & 1221790 & 1222388 & 1222214 \\ \hline		
		\end{tabular}
	\end{center}
	\caption{Features exhibited by the interconnected networks formed in 3 different scenarios, based on the algorithm given in the proof of Theorem \ref{thm:nash}. These results are produced by averaging over 100 instances of random networks and edge costs. The GR network is disconnected in some of the instances, which we indicate with an average diameter of  $\infty$.}
	\label{tab:sim}
\end{table*}

Next, we investigate the impact of heterogeneity in edge costs on the social welfare and the Nash equilibrium networks. 
In the previous scenario we assumed that each player's edge cost was chosen from a uniform distribution on the interval $[0.01, 2500]$. Here, we consider the case where the cost of constructing edges is equal to the mean value of the previous costs, i.e., $c_i=1250$ for $1 \le i \le 500$. Given the  benefit function for the players specified in the previous scenario, all of the players have high edge costs (i.e., $|S_L|=0$). The results are shown in Table~\ref{tab:sim3}.

As the data in Table \ref{tab:sim3} suggests, when players have homogeneous edge costs (equal to the mean of the costs assigned to the players previously), there is a significant decrease in the number of interconnection edges built by the players. Interestingly, this is accompanied by a {\it decrease} in the average distances between the players and their interdependent nodes (in all of the three different scenarios), compared to the situation with heterogeneous edge costs considered previously.  This leads to an {\it increase} in the social welfare under homogeneous edge costs, as shown by the last rows of Tables \ref{tab:sim} and \ref{tab:sim3}.  This phenomenon can be explained via Proposition \ref{prop:high_highcost_vicinity} and the $r$-radius of the players (defined in \eqref{radius}). Specifically, when all players have the same edge cost $c = 1250$, their corresponding $r$-radius is $r_i=2$ for all $1 \le i \le 500$. If a player $P_i$ (associated with node $x_i$) constructs an edge to the center of the sensor network $G_2$, by Proposition \ref{prop:high_highcost_vicinity}, no player $P_j$ with $ d_{G_1}(x_i, x_j) \le 2$ will construct an interconnection edge.  A player $P_j$ with distance $d_{G_1}(x_i,x_j) > 2$ will construct an edge, however.  Thus, the set of players that construct interconnection edges under homogeneous edge costs are spaced apart fairly regularly (i.e., every node is at most distance $2$ in $G_1$ from a node that has constructed an edge).  In the case of heterogeneous edge costs, however, the distances (in $G_1$) between nodes that have constructed edges is no longer bounded by $2$.  Thus, even though there are more nodes that construct edges (due to relatively low edge costs), the average distance between nodes and their interdependent nodes increases.  This also leads to a decrease in social welfare.

\begin{table*}[t]
	\begin{center}
		\begin{tabular}{| l | l | l | l |}
			\hline
			~~~   & SF & ER & GR\\ \hline
			$|E_1|$ & 2970.7 & 2980.8 & 2945.1 \\ \hline
			Diameter of $G_1$ & 4 & 4.42 & $\infty$ \\ \hline
			
			Interconnection Edges Constructed by High Cost Players & 12.7 & 18.8 & 30.8 \\ \hline
			Average Distance Between Interdependent Nodes & 3.95 & 3.92 & 3.87 \\ \hline  
			Social Welfare of the Interconnected Network & 1246825 & 1245300 & 1242250 \\ \hline			
		\end{tabular}
	\end{center}
	\caption{Features exhibited by the interconnected networks formed in 3 different scenarios when all of the players have the same cost $c = 1250$ of constructing edges. These results are produced by averaging over 100 instances of random networks.}
	\label{tab:sim3}
\end{table*}
\section{Related network design problems}
There are many instances of network design problems that have been studied in the computer science and algorithms literature.  
The interconnection network design problem that we investigated in this paper has similarities to the Island-Connection (IC) model that was studied in \cite{island_jackson} where nodes (as network designers) have distance-based utilities. In this model, there are clusters of geographically close nodes (called islands) and it is assumed that the price of intra-island edge construction is less than that of inter-island edge construction. While the IC model considers a homogeneous set of players, the INDG model includes the case that players have different cost and benefit functions. 
Furthermore, the topologies of networks $G_1$ and $G_2$ (which correspond to islands in the IC model) in INDG are fixed, whereas in IC the structure of the islands depends on the cost of intra-island edge construction.
When the cost of intra-island and inter-island edge formation are lower than certain thresholds, \cite{island_jackson} shows that there are complete connections inside the islands. In addition, while the distance between all pairs of inter-island nodes are taken into account in the IC model, our INDG model allows the interdependency network $G_I$ to characterize the set of important inter-island distances.

Another related work is the best response network problem (BRN) \cite{TNSE}, which directly generalizes the classical distance-utility network formation problem given by \eqref{dist_distance_util}.  Specifically, in BRN, there is a central network designer with distance-based utility and a set of pairs of nodes that wish to communicate via short hops.  These pairs are encoded as a network $G_1=(N, E_1)$, where the presence of an edge $(x_1, x_2) \in E_1$ indicates that $x_1$ and $x_2$ wish to be close together in the constructed network.   The utility produced by a constructed network $G=(N, E)$ is
\begin{equation}\label{P_2}
u(G|G_1)=\left( \sum_{(v_i,v_j) \in E_1} b(d_G(v_i,v_j))\right) - c |E|.
\end{equation}
The network that maximizes this utility function is called the best response network to $G_1$. In \cite{TNSE}, we showed that finding a best response network with respect to an arbitrary network $G_1$ is NP-hard. A key difference between BRN and INDG is that in INDG, each node acts as a network designer (i.e., no central network designer) and builds edges to nodes in a different network. In fact, networks $G_1$ and $G_2$ in the definition of the INDG problem have no equivalent correspondence in BRN.

\section{Conclusion}
We introduced the interconnection network design game between two networks $G_1$ and $G_2$. In this game, there is a heterogeneous  (in terms of utility function) set of network designers,  each associated with a node in the network $G_1$. Each node in $G_1$ is dependent on certain nodes in $G_2$, and these dependencies are captured by a network $G_I$. The utility of the players is defined based on the distance-utility function where the objective of each player is to build a set of edges from its associated node to nodes in the network $G_2$ such that distances between its associated node and the nodes it depends on in $G_2$ are minimized. We showed that finding a best response action of a player is NP-hard.   Nevertheless, we showed certain important properties of the best response networks, which enabled us to find a Nash equilibrium for certain instances of the game. Finally, we applied our framework to model the interdependencies between communication and power networks. Our simulations suggest that  the social welfare is larger when players are homogeneous in terms of their edge construction costs, compared to players with heterogeneous edge costs.

One interesting avenue for future research is to consider other classes of utility functions for the players. Another important topic for further research on this problem is to address the scenario where players can build different {\it types} of edges (e.g., representing different types of relationships between the nodes).  Defining appropriate utility functions to capture this scenario, along with a characterization of the resulting Nash equilibria, would be of interest.  Finally, proving existence of Nash equilibria when $G_2$ has an arbitrary structure would be of value.

\bibliographystyle{IEEEtran}
\bibliography{refs}

\end{document}